\documentclass[aps, pra, superscriptaddress, twocolumn, preprintnumbers,amsmath,amssymb,notitlepage, nobalancelastpage,10pt,longbibliography]{revtex4-1}

\usepackage[utf8]{inputenc} 
\usepackage{ amssymb }
\usepackage[T1]{fontenc} 

\usepackage{amsmath}
\usepackage{verbatim}
\usepackage{graphicx}
\usepackage{color}
\usepackage{yfonts}
\usepackage{soul}
\usepackage[normalem]{ulem}
\usepackage{dsfont}
\usepackage{ upgreek }

\newcommand{\be}{\begin{equation}}
\newcommand{\ee}{\end{equation}}
\newcommand{\bea}{\begin{eqnarray}}
\newcommand{\eea}{\end{eqnarray}}

\newcommand\pp{\mathbf{p}}

\newcommand\rr{\mathbf{r}}

\newcommand\xx{\mathbf{x}}

\newcommand{\ev}[1]{\left\langle #1\right \rangle}
\newcommand{\bra}[1]{\left\langle #1\right|}
\newcommand{\ket}[1]{\left| #1\right\rangle}

\newcommand{\braket}[2]{\langle #1|#2\rangle}

\newcommand{\bkev}[3]{\langle #1|#2| #3 \rangle}
\newcommand{\pa}[1]{\left( #1\right)}
\newcommand{\co}[1]{\left[ #1\right]}
\newcommand{\stx}[1]{_\text{#1}}

\newcommand{\fl}[1]{\left\lfloor #1\right\rfloor}

\usepackage[colorlinks=true,citecolor=blue,linkcolor=blue,urlcolor=blue]{hyperref}
\usepackage{subfigure}

\usepackage{color}
\definecolor{darkgreen}{rgb}{0,0.60,.2}

\begin{document}

\title{Analog simulation of high harmonic generation in atoms}
\author{Javier Arg\"uello-Luengo}
\email{javier.arguello@icfo.eu}
\affiliation{ICFO - Institut de Ciencies Fotoniques, The Barcelona Institute of Science and Technology, Av. Carl Friedrich Gauss 3, 08860 Castelldefels (Barcelona), Spain}

\author{Javier Rivera-Dean}
\affiliation{ICFO - Institut de Ciencies Fotoniques, The Barcelona Institute of Science and Technology, Av. Carl Friedrich Gauss 3, 08860 Castelldefels (Barcelona), Spain}

\author{Philipp Stammer}
\affiliation{ICFO - Institut de Ciencies Fotoniques, The Barcelona Institute of Science and Technology, Av. Carl Friedrich Gauss 3, 08860 Castelldefels (Barcelona), Spain}

\author{Andrew S. Maxwell}
\affiliation{Department of Physics and Astronomy, Aarhus University, DK-8000 Aarhus C, Denmark}

\author{David M. Weld}
\affiliation{Department of Physics, University of California, Santa Barbara, California 93106, USA}

\author{Marcelo F. Ciappina}
\affiliation{Department of Physics, Guangdong Technion - Israel Institute of Technology,
241 Daxue Road, Shantou, Guangdong, China, 515063}
\affiliation{Technion - Israel Institute of Technology, Haifa, 32000, Israel}
\affiliation{Guangdong Provincial Key Laboratory of Materials and Technologies for Energy Conversion,
Guangdong Technion - Israel Institute of Technology,
241 Daxue Road, Shantou, Guangdong, China, 515063}

\author{Maciej Lewenstein}
\affiliation{ICFO - Institut de Ciencies Fotoniques, The Barcelona Institute of Science and Technology, Av. Carl Friedrich Gauss 3, 08860 Castelldefels (Barcelona), Spain}
\affiliation{ICREA, Pg. Lluis Companys 23, ES-08010 Barcelona, Spain}

\begin{abstract}
The demanding experimental access to the ultrafast dynamics of materials challenges our understanding of their electronic response to applied strong laser fields. For this purpose, trapped ultracold atoms with highly controllable potentials have become an enabling tool to describe phenomena in a scenario where some effects are more easily accessible and twelve orders of magnitude slower. In this work, we introduce a mapping between the parameters of attoscience platform and atomic cloud simulators, and propose an experimental protocol to access the emission spectrum of high harmonic generation, a regime that has so far been elusive to cold atom simulation. As we illustrate, the benchmark offered by these simulators can provide new insights on the conversion efficiency of extended and short nuclear potentials, as well as the response to applied elliptical polarized fields or ultrashort few-cycle pulses.
\end{abstract}

\maketitle

Over the last three decades, progress in laser technologies has led to significant advances in our ability to manipulate and understand electron dynamics on their natural attosecond ($10^{-18}$~s) timescale~\cite{krauszAttosecond2009, salieres_study_1999, lewensteinPrinciples2009, ciappina_attosecond_2017}. This has triggered the development of a huge range of tools for probing and controlling matter, which includes high harmonic spectroscopy \cite{itataniTomographic2004}, laser-induced electron diffraction \cite{zuo_laserinduced_1996,niikura_sublasercycle_2002}, photoelectron holography \cite{huismans_timeresolved_2011,figueirademorissonfaria_it_2020}, attosecond streaking \cite{hentschel_attosecond_2001,itatani_attosecond_2002}, and reconstruction of attosecond harmonic beating by interference of two-photon transitions \cite{paul_observation_2001,muller_reconstruction_2002}, to name only a small selection. One of the most prominent processes underlying some of these successes is high harmonic generation (HHG), a highly non-linear phenomenon where a system absorbs many photons of the driving laser and emits a single photon of much higher energy.

The experimental realization and interpretation of these complex processes have been guided most often by simplified theoretical descriptions that still capture the main properties of the dynamics. In the field of attoscience, simplifications such as only considering one or two active electrons, disregarding the interaction of the ionized electron with the parent ion during its propagation in the continuum, or performing saddle-point approximations have provided valuable quantitative predictions concerning HHG~\cite{aminiSymphony2019,eberlyHighorder1989,smirnovaMultielectron2013} and other phenomena~\cite{javanainenNumerical1988,paulus_plateau_1994,lewenstein_rings_1995,goreslavskii_electronelectron_2001}. Additional experimental regimes do, however, require a more complete description of the system, including those where multielectronic processes~\cite{smirnovaMultielectron2013}, or Coulomb nuclear potentials play a key role~\cite{popruzhenkoStrong2008} (see Refs.~\cite{popruzhenkoKeldysh2014,figueirademorissonfaria_it_2020} for reviews). This has motivated an intense development of analytical and numerical methods aimed at pushing current computing capabilities.

To circumvent this computation complexity, analog simulation has become an enabling tool to access phenomena with highly controllable devices, whose temporal and spatial scales are more favorable to measure than those naturally present in attosecond physics~\cite{Blatt2012,Bloch2012, Cirac2012}. 
Experimental advances in the engineering of interactions now foster the simulation of quantum chemistry phenomena, such as molecular geometries~\cite{arguello2019analogue,luhmannEmulating2015}, vibronic calculations~\cite{shenQuantum2018,macdonellAnalog2021}, or the presence of conical intersections~\cite{valahuDirect2023,whitlowQuantum2023}. Recently, this experimental benchmark of chemical processes has accessed the response of an atomic system to strong pulses in the regime where the energy imparted by the simulated laser field is strong enough to ionize the target atoms. This is the case of experiments where the incoming field is simulated by the curved shape of a waveguide~\cite{benlevySimulation2023}, or by a shaken potential applied on a neutral atomic species~\cite{senaratneQuantum2018}, following early proposals where the correspondence to the static frame was introduced~\cite{dumWave1998,Arlinghaus2010,Sala2017}. Going beyond the ionization regime, Sala et al.~\cite{Sala2017} noted that controllable Zeeman or Stark shifts can also give access to the relevant regime of HHG in atomic simulators. However, the simulation of the HHG spectrum has remained so far elusive in these experiments where photons or neutral atoms mimic the oscillating electron, as no radiation is emitted by the associated neutrally-charged simulated dipole. 
In this work, we show that current atomic platforms offer a unique opportunity to access and measure the emission spectrum of HHG through absorption measurements. Furthermore, it simulates the physical response of a \emph{single-atom} target. This is in contrast with real experiments, where thousands of atoms are simultaneously driven to collect enough photons to resolve the spectrum, which challenges phase-matching conditions when a large ionization occurs under strong fields. The simulator thus provides an important bridge between experiments and theory, offering controllable experimental access to complex systems that otherwise can only be theoretically approximated.

\begin{figure*}
    \centering
    \includegraphics[width=1.7\columnwidth]{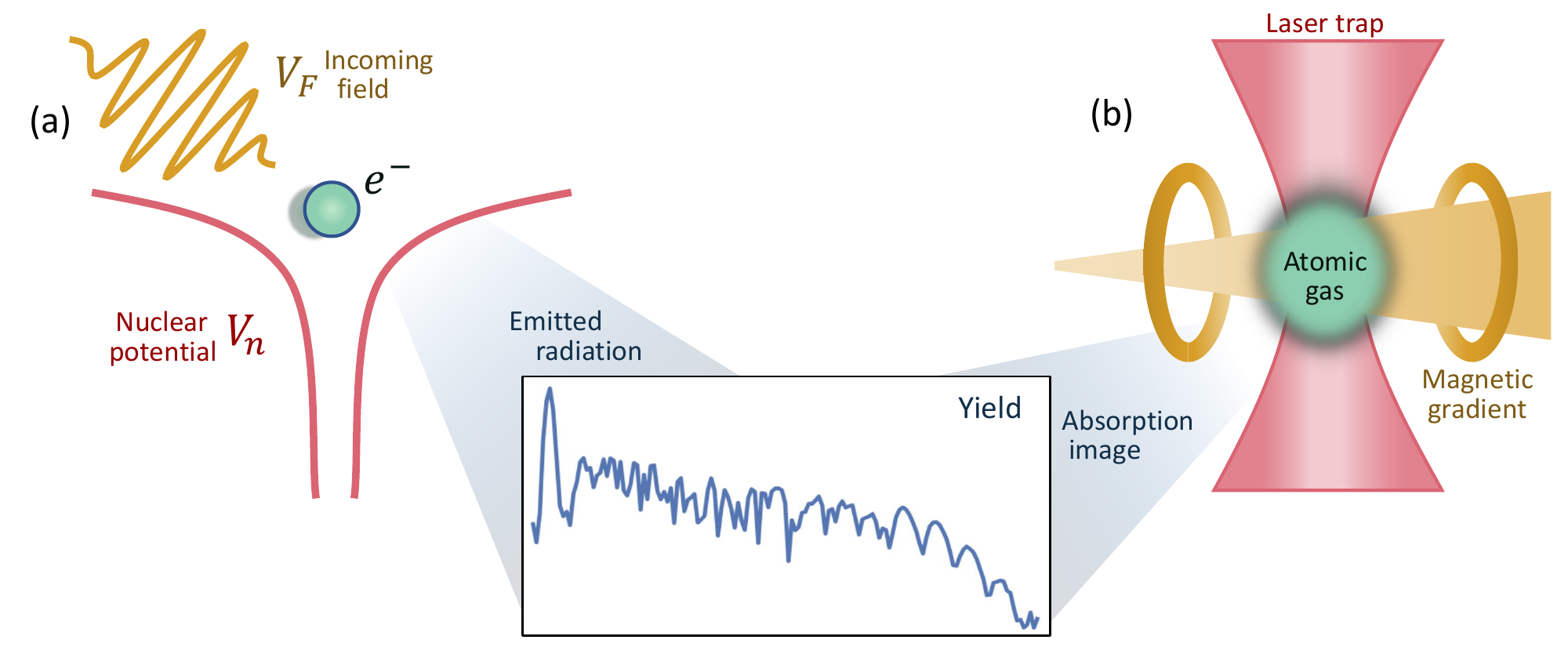}
    \caption{Schematic representation of HHG in atoms (a), and the proposed analog simulator (b). In (a) an ultrafast incoming field (brown) accelerates an electron (green), that is originally trapped by the nuclear Coulomb potential of the atom (red). The resulting oscillation of the charge emits radiation of characteristic harmonic frequencies (inset in blue). The same emission yield can be simulated on the simulator (b), where the characteristic frequencies are retrieved through absorption images of an atomic gas (green) that is trapped by a laser potential (red), and addressed by an external magnetic gradient that is tuned over time (brown). }
    \label{Fig:scheme}
\end{figure*}

The text is structured as follows. First, we present key concepts on attoscience physics and ultracold atom simulators that is of special interest for readers newly exposed to either of these areas. In particular, Sec.~\ref{sec:HHG} discusses the status of current investigation in attosecond science and motivate the different regimes that result from the frequency and strength of the laser field that drives the process. There, we introduce the regime of HHG, and some of its distinctive features. Next, Sec.~\ref{sec:simulation} presents current applications of analog quantum simulation, focusing our attention on the flexibility offered by atomic systems subjected to tunable light fields. There, we introduce an atomic simulator capable of mimicking the dynamics of an electron exposed to a strong oscillatory laser field, and derive a mapping between the experimental parameters of the simulator and the relevant units encountered in attosecond science. We also devise a protocol to access the simulation of the emission yield in HHG, highlighting the range of parameters where this correspondence is valid. As an illustrative example, we show how one can use the simulator to study the effect that short pulses and the ellipticity of the incoming field have on the efficiency of HHG. Sec~\ref{sec:experiment} further presents 
details on the experimental choice of atomic species and laser pulses that are needed to simulate specific targets of common studies in attosecond science, and discuss the main sources of errors that the experimental implementation would encounter, which we numerically benchmark. In Sec.~\ref{sec:discussion}, we present an outlook of the venue for exploration that the proposed simulator opens.

\section{Introduction to HHG}
\label{sec:HHG}
HHG stands as one of the most paradigmatic examples of strong-laser field physics \cite{krauszAttosecond2009,lewensteinPrinciples2009,aminiSymphony2019}. HHG is a highly nonlinear optical process in which a target gets subjected to a very intense ($I \sim10^{13}-10^{14}$ W/cm$^2$), and often short ($\tau \sim 5 - 100$ fs) laser pulse, which typically belongs to the infrared regime ($\lambda \sim700$ nm$ - 5$ $\mu$m). As a result, the oscillating electron emits harmonics that could extend over frequencies hundreds of orders higher than the original driving field. The unique characteristics of HHG, including coherence, ultrashort duration and high intensity, make it an exceptional source of extreme ultraviolet (XUV) radiation \cite{drescherXray2001,silvaSpatiotemporal2015} and, nowadays, it configures the workhorse for the generation of attosecond pulses \cite{krauseHighorder1992,popmintchev_bright_2012}. Consequently, HHG finds applications in various fields such as non-linear XUV optics \cite{kobayashi_27-fs_1998,midorikawa_xuv_2008,chatziathanasiou_generation_2017,tsatrafyllis_ion_2016,bergues_tabletop_2018,nayak_multiple_2018,senfftleben_highly_2020,orfanos_non-linear_2020}, attosecond science \cite{krauszAttosecond2009,aminiSymphony2019}, molecular tomography \cite{itataniTomographic2004} and high-resolution spectroscopy \cite{gohle_frequency_2005,cingoz_direct_2012,silva_high-harmonic_2018,alcala_high-harmonic_2022}.

Although HHG is a process that has been observed in many different targets, such as atoms \cite{lhuillier_high-order_1993,aminiSymphony2019}, molecules \cite{lynga_high-order_1996}, solid-state systems \cite{ghimire_observation_2011,goulielmakis_high_2022} and nanostructures \cite{ciappina_attosecond_2017}, in this work we will focus on simulating HHG process in atoms.
We shall now discuss the conditions required to generate high-order harmonics from atomic targets by introducing relevant parameters.
The first one is the so-called multiquantum parameter $K_0 = I_p/(\hbar \omega)$ \cite{popruzhenkoKeldysh2014}, where $I_p$ is the ionization potential of the corresponding atom, and $\omega$ is the frequency of the driving field. This parameter provides an estimate of the minimal number of  photons required to ionize the electron from the atomic ground state. Here, we are particularly interested in the low-frequency limit, where $K_0 \gg 1$, indicating that ionization requires a significant number of photons. Furthermore, whether a multiphoton absorption process occurs or not also depends on the amplitude $F_0$, of the applied field. This motivates the introduction of the Keldysh parameter $\gamma = \sqrt{I_p/(2U_p)}$ \cite{aminiSymphony2019,popruzhenkoKeldysh2014,keldysh_ionization_1965}, where $U_p = e^2 \xi_0^2/(4m\omega^2)$ is the ponderomotive energy, i.e. the average kinetic energy of an electron with mass $m$ and charge $e$ in the presence of an oscillating laser electric field of amplitude $\xi_0$ [see Fig.~\ref{Fig:scheme}(a)]. 

\begin{figure}
    \centering
    \includegraphics[width=1\columnwidth]{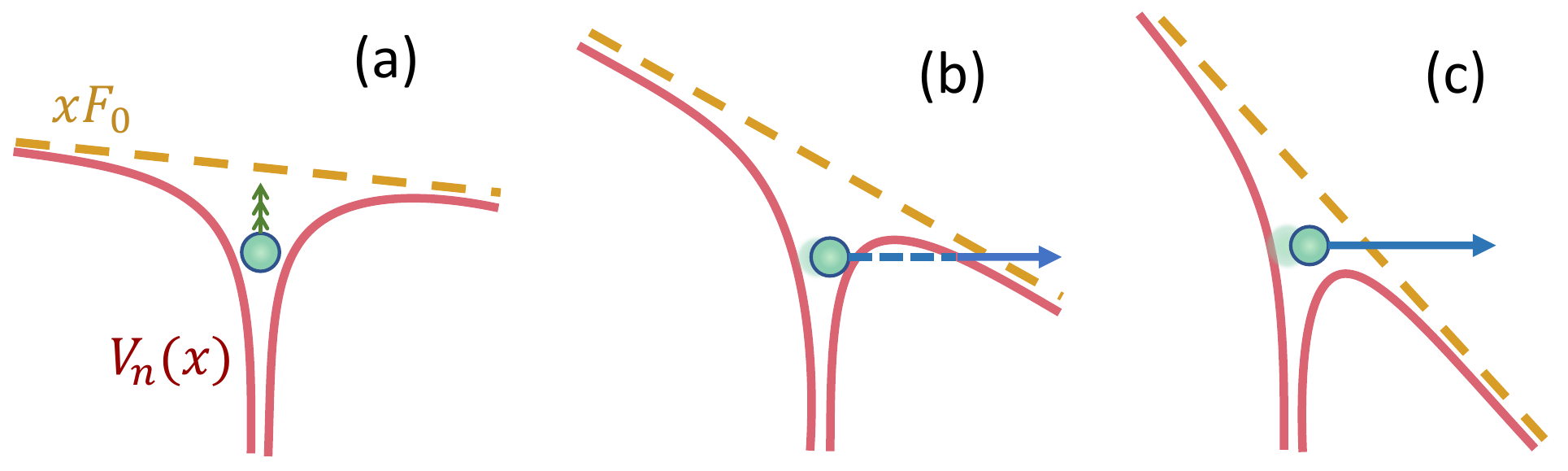}
    \caption{Schematic representation of the different ionization mechanisms at the instant of maximum tilt for different values of field strength. In (a) $\gamma \gg 1$ such that it slightly perturbs the atomic potential, potentially leading to multiphoton ionization mechanisms. In (b) $\gamma \lesssim 1$, such that the field is strong enough to allow for tunneling ionization. In (c), the electric field has reached a critical value that makes the maximum of the barrier to coincide with the energy level of the electron, leading to an over-the-barrier ionization mechanism. Note that in all these cases, we are working in the regime $K_0 \gg 1$.}
    \label{Fig:HHG:section}
\end{figure}

In the case $K_0 \gg 1$, the multiphoton regime is observed when $\gamma \gg 1$, indicating that the field only slightly perturbs the atomic potential [see Fig.~\ref{Fig:HHG:section}~(a)]. The regime of interest for HHG processes corresponds to $\gamma \lesssim 1$, known as the \emph{tunneling regime}, where the external field force is comparable to the atomic potential, which gets distorted and forms an effective potential barrier through which the electron can tunnel out [see Fig.~\ref{Fig:HHG:section}~(b)]. Finally, and for the sake of completeness, we have the regime of over-the-barrier ionization that, typically~\cite{mulserIntense2010}, happens when the electric field reaches a critical value that makes the barrier maximum to coincide with the energy level of the electron ground state [see Fig.~\ref{Fig:HHG:section}~(c)].

Within the tunnelling regime, the three-step model, also referred to as the \emph{simple-man's model} \cite{krauseHighorder1992,corkum_plasma_1993,kulander_dynamics_1993}, provides a powerful picture of the underlying electron dynamics behind HHG. The steps within this model are as follows: the electron (i) tunnels out from the parent atom through the barrier formed by the Coulomb potential combined with the dipole interaction of the field, (ii) oscillates in the continuum under the influence of the laser electric field and, if it passes around the nucleus' vicinity, (iii) can recombine back to the ground state emitting harmonic radiation. The energy of the emitted radiation upon recombination depends on the electronic kinetic energy at the moment of recombination and the ionization potential of the atom. However, the maximum kinetic energy that the electron can acquire from the field during its propagation is limited, leading to a natural cutoff frequency $\omega_c$ for the highest harmonic order in HHG, which is determined by $\hbar \omega_{c} = I_p + 3.17 U_p$. The nonlinear character of the HHG process can be understood from its main features, namely (i) a strong decrease in the low-order harmonics amplitude, (ii) a plateau, where the harmonic yield is almost constant, and (iii) a sudden cutoff, given by the above presented classical formula~\cite{krauseHighorder1992}.

Based on the previous discussion, it is evident that the polarization character of the driving field can have significant consequences on the generated harmonic radiation. For instance, when an elliptically polarized driver is considered, the ionized electron may miss the parent ion, resulting in the absence of the recombination event \cite{corkum_plasma_1993}. This phenomenon has been extensively studied in both experimental \cite{budilInfluence1993,burnett_ellipticity_1995,weihe_polarization_1995,weihe_measurement_1996,antoine_polarization_1997,schulze_polarization_1998} and theoretical \cite{becker_effects_1994,antoine_theory_1996} works, demonstrating a reduced HHG conversion efficiency, i.e.~the ratio between the outgoing and incoming photon fluxes, as the driving laser ellipticity increases. As an alternative strategy, a combination of two drivers that have different ellipticities and frequencies can be used to generate bright phase-matched circularly-polarized high harmonics, as shown in Refs.~\cite{fleischerSpin2014, kfirGeneration2015, milosevicGeneration2000, pisantySpin2014,pisantyalatorreElectron2016}.

While we will focus on HHG, strong laser-matter interactions can also give rise to other fascinating phenomena, including Above-Threshold Ionization (ATI) \cite{agostini_free-free_1979,delone_multiphoton_2000,milosevicAbovethreshold2006,lewensteinPrinciples2009,agostini_chapter_2012} and Non-Sequential Double Ionization (NSDI) \cite{lhuillier_multiply_1983,corkum_plasma_1993,walker_precision_1994,feuerstein_separation_2001,faria_electron_2011,becker_theories_2012}. In ATI, an electron is ionized by the strong-laser field, surpassing the ionization threshold of the corresponding atom by absorbing more photons than the ones required for ionization. The typical observable measured in ATI is the photoelectron spectrum, which exhibits distinctive peaks at electron kinetic energies  separated by the energy of a single photon of the driving field \cite{agostini_free-free_1979,milosevicAbovethreshold2006}. In the non-perturbative (tunneling) regime, these peaks form a plateau that extends over electron energies on the order of $10\, U_p$ \cite{hansch_resonant_1997,milosevicAbovethreshold2006}. On the other hand, NSDI occurs when an ionized electron undergoes rescattering with its parent ion, resulting in the ionization of a second electron. This phenomenon is reflected in the ion yield, which exhibits a \emph{knee} structure at a specific intensity threshold. The presence of this distinctive feature signifies a sudden change in the energy distribution of the emitted electron pairs, indicating a transition from sequential to non-sequential ionization processes~\cite{lhuillier_multiply_1983,walker_precision_1994}. 

Here, we demonstrate the capability of analog simulators to accurately replicate the key characteristics of the HHG processes in atoms. Specifically, it recovers the main features of a typical HHG spectrum --a plateau extending for few harmonic orders followed by a cutoff-- with an harmonic yield that reduces for increasing values of the ellipticity. We study how the conversion efficiency of the harmonics depends on the values of $K_0$ and $\gamma$, and discuss how the simulated spectrum can be measured in practice for analog simulators, also providing estimates on relevant quantities towards feasible experimental implementations.

\section{Analog quantum simulation}
\label{sec:simulation}
The numerical simulation of chemical problems generally requires to describe many electrons that interact with external fields, the nuclei, and among themselves through Coulomb interactions. Even if one considers the nuclear positions fixed due to their larger mass (the Born-Oppenheimer approximation \cite{born_zur_1927,atkins_book}), this is an extremely challenging task, as the associated Hilbert space grows exponentially with the number of electrons.

Over the last few decades, an alternative route to study electronic problems has emerged, based on using quantum devices that can better capture the complexity of the system. This idea was first proposed by Feynman as a way of preventing the exponential explosion of resources of quantum many-body problems~\cite{Feynman1982}, and later formalized by Lloyd~\cite{Lloyd1996}.  
Complementary to current efforts in digital simulation~\cite{arrazolaQuantum2021, aruteQuantum2019, googleaiquantumandcollaboratorsHartreeFock2020a} (where the problem is first encoded as qubits and gates addressed by a general-purpose quantum device), simulators based on ultracold atoms have become an enabling tool, already addressing quantum matter phenomena that the most advanced classical computers cannot compute~\cite{choi16a,trotzky12a}. 

At low temperatures, the interaction among atoms can be highly engineered with external lasers, which allows one to induce a rich variety of effective Hamiltonians on a highly controllable platform~\cite{Cirac2012,daleyPractical2022}.
Early experiments dealt with condensed matter problems~\cite{Greiner2002,Jordens2008,Jaksch1998}, detecting a transition between the superfluid and Mott insulating phases of effective Hubbard models. More recently, atomic simulators have experimentally addressed problems related to high-energy physics, such as Gauge theories, both in lattice geometries~\cite{schweizerFloquet2019,gorgRealization2019,aidelsburgerCold2022} and the continuum~\cite{frolianRealizing2022}. 
An exciting perspective consists of extending the benefits of analog quantum simulation to the field of chemistry and the response of atoms and molecules to external fields. Soon after the first experimental realization of bosonic gases~\cite{andersonObservation1995, davisBoseEinstein1995}, the mapping between degenerate atomic gases and single-electron dynamics was noticed~\cite{dumWave1998,esryHartreeFock1997}. As compared to a real material, where many target atoms are present, here the simulated electron moves in a cleanly isolated environment. Furthermore, the typical energy-scales of these experiments are in the range of kHz-MHz, which provides a temporal magnification of the simulator, where attosecond pulses are associated to convenient timescales of $\mu$s-ms, i.e. $10^{9}\sim 10^{12}$ times slower. The additional tunability and accessibility of these simulators can thus offer a complementary tool to investigate ultrafast phenomena.

Using the Kramers-Henneberger correspondence, the shaking of the optical trap can mimic the effect of an external force~\cite{ Sala2017, rajagopalQuantum2017,Arlinghaus2010}, which has allowed for the recent experimental simulation of ATI processes using a bosonic gas of $^{84}$Sr~\cite{senaratneQuantum2018}, where the kinetic energy of the simulated ionized electrons is accessed through a time-of-flight measurement (TOFM) of the atoms emitted during the shaking. Extending this approach to the HHG regime is however challenging, as the needed inertial force corresponds to strong nuclear potentials (see Appendix A), and the associated photonic spectrum is neither emitted during the oscillation of neutral atoms, nor directly revealed by the TOFM.
The present contribution advances the study of HHG simulation in different directions. In particular, (i) we establish a correspondence between the experimental parameters in the simulator and the $K_0$ and $\gamma$ parameters conventionally used in attosecond science, presenting specific configurations associated with standard choices of atomic targets and ionizing pulses; (ii) we show that the region of maximum HHG conversion efficiency can be accessed in atomic simulator platforms where the external pulses are simulated by existing Stark and Zeeman shifts; (iii) we introduce a scheme to experimentally access the generated harmonic spectra through absorption measurements of the atomic gas, and characterize the main sources of error.

\subsection*{The simulator}
Whenever multielectron processes can be neglected, the underlying physics of strong field processes can be described by a single-active electron. The dynamics of this electron is dictated by the Hamiltonian
\begin{equation}
\label{eq:hamiltonian}
    \hat H(\rr,t)=\frac{\pp^2}{2m}+V_n(\rr)+V_F(\rr,t)\,,
\end{equation}
which accounts for its kinetic energy, the attractive nuclear potential, and the incident laser field, respectively.
Following the dipole approximation, the interaction with a laser field linearly polarized along the $\xx$ axis writes as, 
$V_F(\rr,t)=F(t) x$,
where $F(t)=F_0 f(t) \sin\pa{\omega t+\varphi}$ and $f(t)=\sin^2\co{\omega t/(2 n_c)}$ is the carrier-envelope for a pulse with $n_c$ cycles and wave phase $\varphi$. Here, $F_0=e\xi_0$ is the maximum force imparted by the field on the electron.

In this proposal, the dynamics of the electron inside the nuclear potential is simulated by an atomic Bose gas optically trapped by a laser beam, whose spatial profile can be highly engineered, even dynamically, with the use of spatial light modulators~\cite{boyerDynamic2006, mcgloinApplications2003} or digital mirror devices~\cite{muldoonControl2012, renTailoring2015}. Working in the single-active electron approximation, this can allow one to engineer an effective potential for the valence electron based on an \emph{ab initio} calculation of the multielectronic species, which can even provide quantitative agreement on the emitted spectrum, as theoretically studied in Refs.~\cite{satoTimedependent2016,reiffSingleactive2020}. As a first experimental step for the simulator, here we will consider the natural Gaussian profile of a laser beam of waist $r_0$ and depth $V_0$ in a one-dimensional system
\begin{equation}
\label{eq:nuclearPot}
V_n(x,r_0)=-V_0 \exp\left[-x^2/(2r_0^2)\right].
\end{equation}
Similarly to the widely used atomic units, it becomes convenient to define the natural units of this system as $[\hbar]=[m]=[V_0 r_0]=1$. In these units (that we denote as $[\cdot]$ along the text) the nuclear potential is fully characterized by the width of the trap, as it follows from the relation $[V_0]=[r_0]^{-1}$, and approximates a quantum harmonic oscillator whenever the associated zero-point motion of the oscillator is smaller than the waist of the beam, $[r_0]\gg1$. 

\begin{figure}[tbp]
\centering  \includegraphics[width=\linewidth]{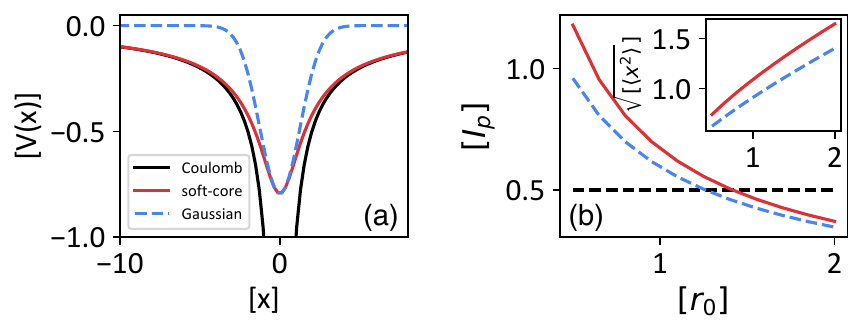}
      \caption{ (a) Comparison between the Coulomb potential, $-1/|x|$ (black line), the soft-core one, $V\stx{sc}(x,r_0)$ (red) and the Gaussian potential $V_n(x,r_0)$ (blue) for $[r_0]=1$. (b) Ionization energy associated to the Gaussian and soft-core potential (same colour code as before), for different values of $r_0$. Black dashed line follows the Hydrogen-like condition, $[I_p]=0.5$. Inset indicates the average ground-state width. See Appendix C for details on the numerical calculation.}
 \label{fig:pote_Ip}
\end{figure}

In Fig.~\ref{fig:pote_Ip}(a), one can see that the simulated Gaussian potential (dashed blue line) matches at short distances ($x/r_0\ll1$) the widely used soft-core nuclear potential of the form, $V\stx{sc}(x,r_0)=-1/\sqrt{x^2+r_0^2}$  (red line, see Appendix B for further details)~\cite{javanainenNumerical1988, suModel1991}. To highlight this connection between the simulator and the soft-core potential, in Fig.~\ref{fig:pote_Ip}(b) we calculate the ionization energy associated with both nuclear potentials for different values of $r_0$, observing discrepancies smaller than $10\%$ for $[r_0]>1$. Focusing on the average width, $\sqrt{\ev{x^2}}$, of the ground-state, the long-range polynomial scaling of the soft-core potential leads to more extended eigenstates than the short-range Gaussian trap~\cite{collinsModel1988, geltmanIonization1977} (see inset), which can be experimentally explored by further shaping the laser beam. As compared to a real target with a fixed nuclear potential, the tunability offered by the simulator allows one to benchmark how the predictive power of conventional numerical methods is influenced by the range of the studied nuclear field. For example, this is relevant for the strong field approximation, where the Coulomb potential along the excursion time of the electron is often numerically disregarded~\cite{aminiSymphony2019}.

To mimic the effect of an incoming electric field under the dipole approximation, the atomic cloud is subjected to a time-dependent linear energy gradient, $V_F(\rr,t)$, which can be created by an optical Stark shift that is proportional to the intensity of an applied off-resonant laser field. In current platforms, a linearly-varying intensity can be created with a spatial light-modulator~\cite{choi16a}, an acousto-optical device, or simply by using the slope of a Gaussian beam whose intensity and position can be dynamically adjusted. For atomic levels that are sensitive to magnetic fields, one can alternatively rely on Zeeman shifts induced by linear magnetic field gradients created and modulated with current-carrying coils~\cite{fancherMicrowave2018,linSynthetic2009, maPhotonAssisted2011}, as represented in Fig.~\ref{Fig:scheme}(b).

\subsection*{Simulation of HHG emission yield}
\label{sec:emissionYield}
While there is not an analog equivalent to the photons emitted during HHG, its emission yield can be accessed through the time-dependent dipole moment $ d(t)=\bra{\psi(t)}x\ket{\psi(t)}$, or its associated time-dependent dipole acceleration, $d_a(t)=\bra{\psi(t)}\ddot x\ket{\psi(t)}$, where $\ket{\psi(t)}$ denotes the atomic state at time $t$. In HHG experiments, the spectrum of energies for photons emitted over the duration of the laser pulse is characterized by the Fourier transform, $d_a(\Omega)=\int dt \, d_a(t) e^{-i\Omega t}$~\cite{millerTime2016}.

\begin{figure}[tbp!]
\centering  \includegraphics[width=\linewidth]{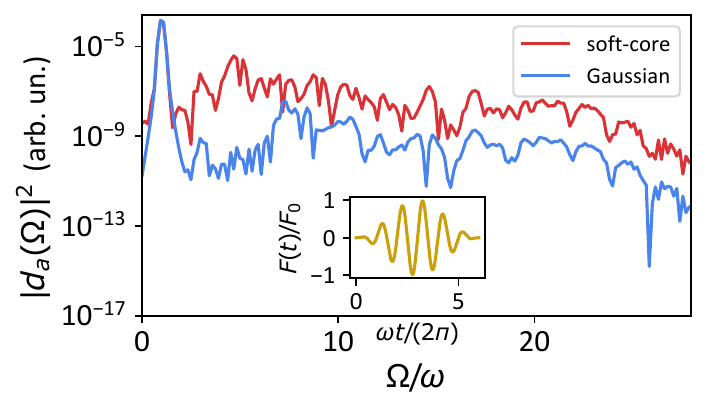}
      \caption{Simulated emission yield, $d_a(\Omega)$, associated to a Gaussian (blue) and soft-core potential (red line). (Inset) Sinusoidal pulse, $F(t)=F_0\sin^2(\omega t/2n_c)\sin(\omega t)$.  Parameters: $[F_0]=0.053$, $[\omega]=0.057$, $[r_0]=1.26$, and $n_c=6$. See Appendix C for details on the numerical calculation.}
 \label{fig:HHGpulse_field}
\end{figure}

In Fig.~\ref{fig:HHGpulse_field} we use a time-dependent Schr\"odinger equation (TDSE) to calculate the dipole acceleration $d_a(\Omega)$ in one spatial dimension. From Fig.~\ref{fig:pote_Ip}(b), we choose parameters compatible with the Hydrogen ionization potential $[r_0]=1.26$, so that the corresponding dissociation energy is $[I_p]=0.5$, and a 6-cycle pulse with an associated laser field of wavelength 800 nm and intensity $10^{14}$ W/cm$^2$. There, we see that the short-range correspondence between the soft-core (red line) and Gaussian potentials (blue) translate into a qualitative agreement of the resulting harmonic spectrum. 

To access this quantity in the simulator, one option consists of measuring the atomic spatial density and following the Ehrenfest theorem~\cite{gordonQuantitative2005}
\begin{equation}
    d_a(t)=-\bra{\psi(t)}\nabla V_n(x) \ket{\psi(t)}\,,
\end{equation}
where $\nabla V_n(x,r_0) $ denotes the gradient of the trapping potential. For the Gaussian trap expressed in Eq.~\eqref{eq:nuclearPot}, this gradient can be approximated as,
$    \nabla_x V_n(x,r_0)\propto V_n(x+r_0/\sqrt{2},r_0/\sqrt{2})-V_n(x-r_0/\sqrt{2},r_0/\sqrt{2})\,.$
Therefore, an absorption measurement allows one to access $d_a(t)$ by quantifying the asymmetry in the atomic population for positions at a characteristic separation $r_0/\sqrt{2}$ from the center of the trap. By repeating the measurement at different times, one can recover the emission spectrum by Fourier transforming $d_a(t)$. 

As an alternative, one can also access the time-dependent dipole velocity, $d_v(t)=\bra{\psi(t)}\dot x\ket{\psi(t)}\,,$ whose spectra is related to the emission yield as, $d_v(\Omega)=id_a(\Omega)/\Omega$, in the case of finite pulses~\cite{baggesenDipole2011}. Interestingly, the velocity components of the atomic cloud at a given time can be conveniently accessed in the simulator through a TOFM, where the nuclear trap is suddenly released and the gas expands ballistically. Once the gas has expanded for a time $\tau$ beyond the initial size of the cloud, the velocity component, $v$, of the state of interest is associated to an absorption detection of the cloud at a distance $x=v\tau$ from the initial trap. Given that these separations are much larger that the size of the original cloud, this approach improves the accuracy of the reconstructed emission yield for a given spatial resolution.

\begin{figure}[tbp!]
\centering
\includegraphics[width=0.9\linewidth]{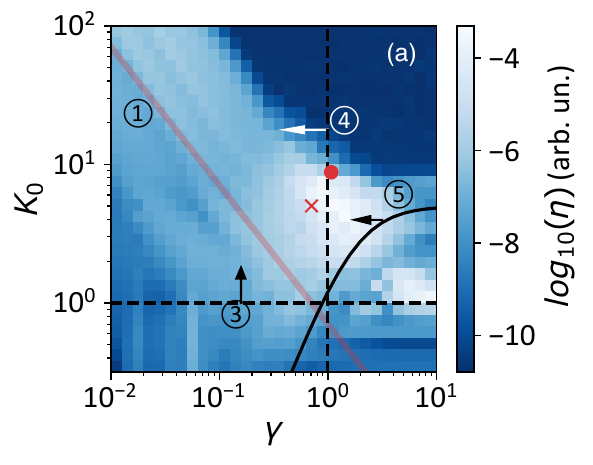}
\includegraphics[width=1\linewidth]{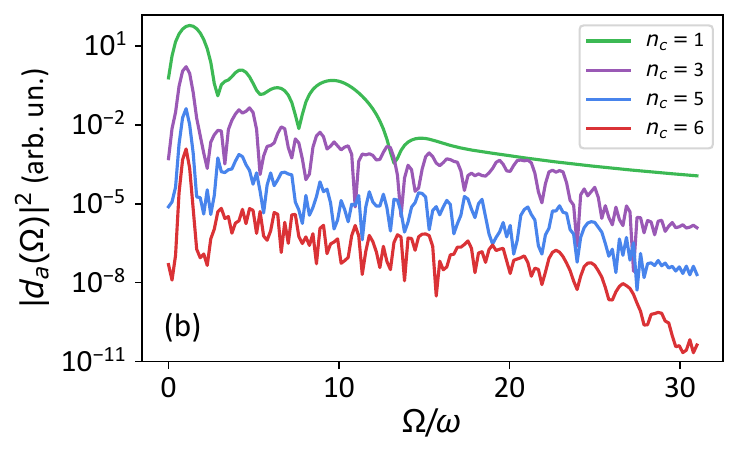}
      \caption{(a) Conversion efficiency, $\eta$, of the 5th harmonic for the Gaussian potential in Fig.~\ref{fig:HHGpulse_field}, a six-cycle pulse ($n_c=6$), and different values of Keldysh ($\gamma$) and multiquantum parameter ($K_0$). Red line \textcircled{1} follows the region where a higher efficiency is expected from Eq.~\eqref{eq:tunnelCrit}, dashed lines \textcircled{3} and \textcircled{4} describe the HHG region ($K_0\geq 1$ and $\gamma\leq 1$, respectively). \textcircled{5} follows the frontier where the selected harmonic is below the cutoff frequency in Eq.~\ref{eq:OmegaCrit}. The dipole approximation \textcircled{2}, associated to the upper bound in Eq.~\eqref{eq:dipole}, lies above the represented range of values for $K_0$. Arrows point towards the region of validity for each of these conditions. Circled red marker indicates the Gaussian configuration in Fig.~\ref{fig:HHGpulse_field}, while the crossed marker indicates the point $K_0=6.31$ and $\gamma=0.81$, associated to $[F_0]=0.171$ and $[\omega]=0.108$. The simulated emission spectrum of the latter configuration is illustrated in panel (b) for incoming pulses with a number of cycles running from $n_c=6$ (red) to $n_c=1$ (green). To improve visibility, an arbitrary vertical shift between the curves has been applied.}
 \label{fig:gamK}
\end{figure}

In Fig.~\ref{fig:gamK}(a) we calculate the conversion efficiency~\cite{gkortsasScaling2011}
\begin{equation}
    \eta=\int_{\Omega-1}^{\Omega+1} |d_a(\omega)|^2d\omega/|F_0|^2\,,
\end{equation}
for one of the harmonics in the plateau region, a fixed nuclear potential $[I_p]=0.5$, and different values of $[\omega]=[I_p]/K_0$ and $[F_0]=\sqrt{2[I_p]^3}/(K_0 \gamma)$. 
In order to interpret the observed regimes, both $K_0$ and $\gamma$ describe a complete set of parameters that characterizes the response of the simulated system to the oscillatory field. For example, expressed in natural units, the last harmonic $\Omega_c=\omega\stx{c}/\omega$ that becomes accessible below the cutoff energy writes as
 \begin{equation}
 \label{eq:OmegaCrit}
     \Omega_c=K_0\pa{1+3.17/\gamma^2} \,. 
 \end{equation}
 As expected, a higher yield appears above the continuous line \textcircled{5} where the condition, $\Omega\leq \Omega_c$, is satisfied. 

To enhance the conversion efficiency, it is also desirable not to be too deep in the tunneling regime. 
An optimal situation occurs when the maximum tilt of the nuclear potential, $F_0 r_0$, is comparable to $I_p$
\begin{equation}
\label{eq:tunnelCrit}
    \gamma K_0 \approx   [r_0] \sqrt{[I_p]}, 
\end{equation}
and we observe that the region of the largest conversion efficiency follows this heuristic scaling (red line \textcircled{1}).

Regarding the mapping to attoscience, the dipole approximation followed in Eq.~\eqref{eq:hamiltonian} requires that the magnetic component of the incoming field can be disregarded. This imposes an upper bound \textcircled{2} on the intensity of the field, which writes as~\cite{reissLimits2008}
\begin{equation}
\label{eq:dipole}
    K_0 /4 \ll \gamma^2[c] \,.
\end{equation}
 Preventing relativistic velocities on the accelerated particle ($U_p\ll mc^2$) also induces a lower bound on the Keldysh parameter
\begin{equation}
\label{eq:relativistic}
    \sqrt{[I_p]} \ll \gamma [c] \,, 
\end{equation}
which is less demanding than the dipole approximation along this region of greatest conversion efficiency in Eq.~\eqref{eq:tunnelCrit} for $\gamma\ll 1$. In the natural units of the simulator, one can see that $[c]\sim 10^{11}$ for the parameters studied in Fig.~\ref{fig:gamK}(a), which places this upper bound above the represented range of values for $K_0$.

One should note that the harmonic radiation measured in in actual attoscience HHG experiment results from a collective phenomena, involving $10^{10}\sim 10^{12}$ atoms that emit coherent radiation \textit{in phase}. When comparing theory and experiment, it is therefore mandatory to include macroscopic propagation effects, which is a formidable computational task that is only addressed by a limited number of models~\cite{gaardeMacroscopic2008}. In this simulator, however, all atoms in the bosonic gas contribute to magnify the effects manifested by a simulated single electron. The resulting harmonic spectrum is thus unaffected by the additional phase-matching condition that is often encountered in attosecond science experiments, providing clean access to the simulated single-atom dipole acceleration. Furthermore, the more favorable energy, spatial and temporal scales in the simulator offer a benchmark that is less affected by the energy resolution, the uncertainty in the laser intensity and duration, or the limited dynamic range of spectroscopic measurements in ultrafast experiments~\cite{mulserIntense2010}.

\begin{figure}[tbp!]
\centering
\includegraphics[width=0.9\linewidth]{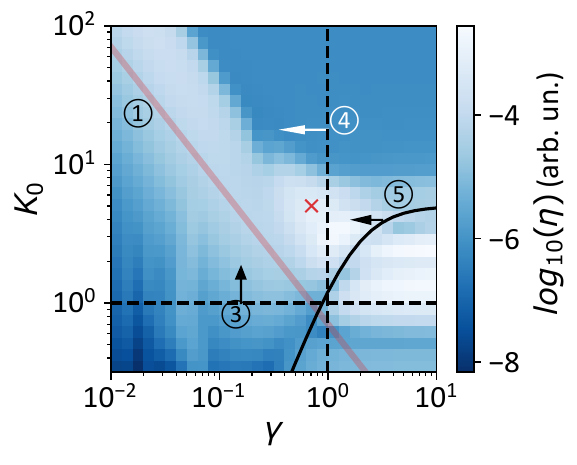}
      \caption{Conversion efficiency, $\eta$, of the 5th harmonic for the Gaussian potential in Fig.~\ref{fig:HHGpulse_field}, a single-cycle pulse, $n_c=1$, and different values of Keldysh ($\gamma$) and multiquantum parameter ($K_0$). Red line \textcircled{1} follows the region where a higher efficiency is expected from Eq.~\eqref{eq:tunnelCrit}, dashed lines \textcircled{3} and \textcircled{4} describe the HHG region ($K_0\geq 1$ and $\gamma\leq 1$, respectively). \textcircled{5} follows the frontier where the selected harmonic is below the cutoff frequency in Eq.~\ref{eq:OmegaCrit}. Arrows point towards the region of validity for each of these conditions. The crossed marker indicates the point $K_0=6.31$ and $\gamma=0.81$, whose emission spectrum is depicted in green in Fig.~\ref{fig:gamK}(b).}
 \label{fig:gamK_nc1}
\end{figure}

The high tunability of the induced incoming pulse is also one of the advantages of the simulator, as compared with the fast high-intensity lasers that are typically needed in attoscience. For example, this allows one to easily simulate the response of the system to ultrashort pulses, a configuration that is otherwise demanding to access in real attosecond experiments.
In Fig.~\ref{fig:gamK}(b), we show the simulated emission yield for pulses with different number of cycles. As the pulses get shorter, we observe that the plateau structure vanishes and the last harmonics disappear, even though the theoretical cutoff frequency $\omega_c$ only depends on the ionization energy and ponderomotive energy, which are the same for all the curves. Intuitively, fewer interference processes can take place when the number of cycles of the pulse decreases, which reduces the number of harmonics that are visible on the emission spectrum as one approaches $n_c=1$ (green line).
Focusing on this latter case, in Fig.~\ref{fig:gamK_nc1} we simulate the response of the system to a single-cycle pulse ($n_c=1$), for different values of $K_0$ and $\gamma$. When focusing on one of the lowest harmonics (the 5th one), we observe that the region of largest emission still remains well described by the different conditions introduced in Fig.~\ref{fig:gamK}(a). 

\begin{figure}[tbp!]
\centering
  \includegraphics[width=\linewidth]{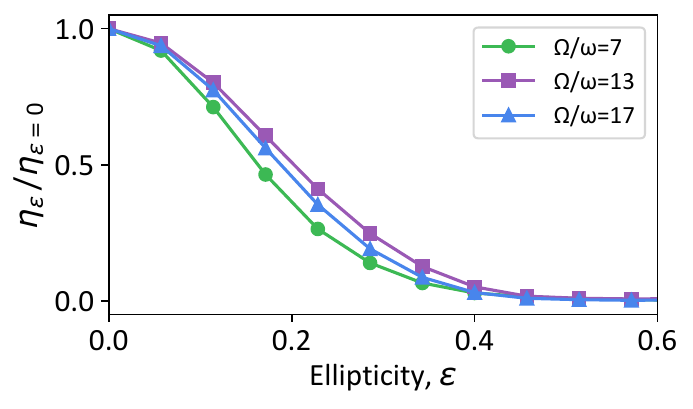}
      \caption{Conversion efficiency of different harmonics (see legend) in a 2D simulator along the x-axis, as a function of the ellipticity of the incoming field. For each harmonic, the efficiency is normalized by the value at $\varepsilon=0$. \emph{Parameters} as in Fig.~\ref{fig:HHGpulse_field}, now with a 2D Gaussian nuclear potential. }
 \label{fig:HHGpulse_elliptic}
\end{figure}

In addition to the linearly polarized fields considered so far, the simulator also allows one to induce oscillations in a second axis, $F_{x(y)}=F_{0x(0y)}f(t)\sin\co{\omega t+\varphi+(\pi/2)},$ when we extend the systems to two dimensions. By controlling the ratio between the two amplitudes, i.e. the ellipticity $\varepsilon=F_{0y}/F_{0x}$, one can induce elliptic [$\varepsilon\in(0,1)$] and circularly-polarized fields ($\varepsilon=1$).
In HHG, the introduction of ellipticity in the laser beam leads to the deflection of the returning electron, causing it to deviate from its intended path toward the parent nucleus. This results in a decrease of the overlap between the wavepackets of the returning electron and the initial bound state, as it has been experimentally observed~\cite{budilInfluence1993}. In Fig.~\ref{fig:HHGpulse_elliptic} we calculate the change in the conversion efficiency of different harmonics as we change from a linear field ($\varepsilon=0$) to an elliptically-polarized field. As expected, we observe a decline in the intensity of harmonics as the ellipticity of the laser beam increases, which is more pronounced for the lowest harmonics. 

\begin{figure}[tbp!]
\centering
  \includegraphics[width=\linewidth]{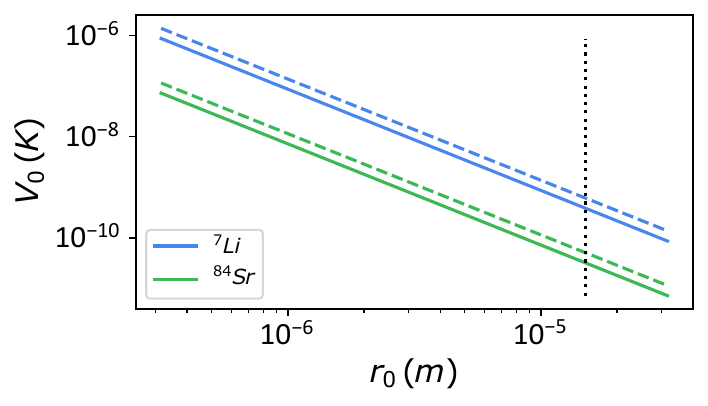}
      \caption{  Nuclear potential strength, $V_0$ as a function of the width of the beam, $r_0$, following Eq.~\eqref{eq:Vexp}. Green and blue continuous lines correspond to $^{84}$Sr and $^7$Li, respectively, for $[I_p]=0.5$ (continuous line) and $[I_p]=0.19$ (dashed line). Dotted line shows the condition $r_0=15\,\mu m$ of Ref.~\cite{senaratneQuantum2018}.}
 \label{fig:HHGpulse_temperature}
\end{figure}

\section{Experimental implementation}
\label{sec:experiment}
At this point, it is worth exploring the feasibility of the experimental parameters associated with this implementation.
For a fixed geometry $[r_0]=1.26$ (associated to $[I_p]=0.5$) the needed nuclear potential scales as 
\begin{equation}
\label{eq:Vexp}
   V_0=\frac{\hbar^2}{m r_0^2}[r_0]\,. 
\end{equation}
In the case of $^{84}$Sr and a Gaussian beam with waist $r_0=1\,\mu$m, the configuration $K_0=6.31$ and $\gamma=0.81$ [illustrated in Fig.~\ref{fig:gamK}(b)] corresponds to parameters, $F_0= 1.55\cdot 10^{-26}\,$N, $\omega =94.7\, \text{Hz}$, and $V_0=  7.22\, \text{nK}\cdot k_B$.
Using moderate conditions of 1 Watt of 532 nm light shaped to give a linear intensity gradient across a 100 $\mu$m $\times$ 100 $\mu$m area, it is possible to achieve the needed values of $F_0$ for $^{84}$Sr atoms, and even produce forces two orders of magnitude stronger. However, one can observe that the associated trap depth $V_0\sim 10\,$nK, is well below the trap depth of around  $10\, \mu$K used in previous attoscience simulators with Sr~\cite{senaratneQuantum2018}. 

The trap depth sets an upper limit to the temperature of atomic clouds that can be trapped by the laser potential. To increase the range of allowed temperatures for the experiment, Eq.~\eqref{eq:Vexp} indicates that more relaxed cooling conditions would benefit from choosing narrower beams and lighter atomic species. This is the case of $^7$Li and $r_0\sim 1\,\mu$m, where the parameters needed to simulate the same attoscience configuration are: $F_0= 1.86\cdot 10^{-25}\,$N, $\omega =1.14\, \text{kHz},$ and $V_0\sim 0.1\, \mu \text{K}\cdot k_B$.
The associated critical temperature is then consistent with state-of-the-art experiments~\cite{pollackExtreme2009a}.
As mentioned in the previous section, for magnetic ground states, the oscillatory force can be applied by a tunable magnetic field gradient. In the case of the $|m_F|=2$ hyperfine state of $^7$Li, easily attainable gradients of 50 G/cm translate into the needed range of forces $F_0$, or even one order of magnitude larger.

In Fig.~\ref{fig:HHGpulse_temperature}, we show the scaling of the trap depth with the beam waist for $^{84}$Sr (green) and $^7$Li (blue lines). We also show that more relaxed cooling conditions would appear in the simulation of weaker ionization energies, as we illustrate with dashed lines for the Sodium first ionization energy, $[I_p]=0.19$. Depending on the choice of atomic species and width of the trap, we observe a range of 3 orders of magnitude on the associated trap depth, which illustrates the flexibility that these simulators offer to access the regime of interest.

\subsection*{Further experimental considerations}
\begin{figure}[h!]
\centering
  \includegraphics[width=\linewidth]{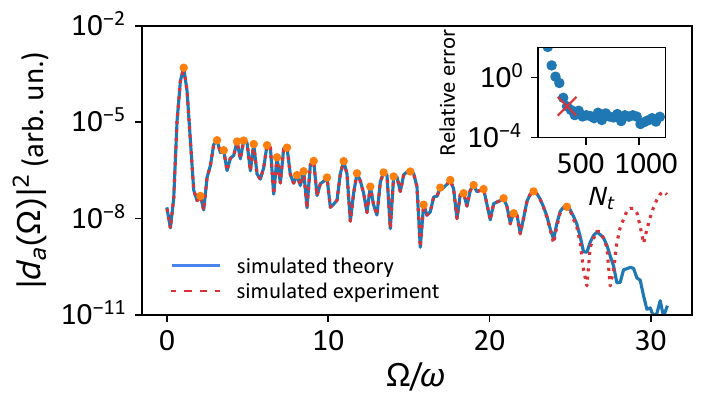}
      \caption{Emission yield extracted from the discrete Fourier transform of $N_t=300$ temporal intervals uniformly distributed along the duration of the pulse (red dotted line), as compared to the converged spectrum ($N_t=3000$, blue line). Orange dots indicate the positions of local maxima before the cutoff energy. In the inset, we show the relative error of these local maxima for increasing number of temporal divisions, $N_t$, where the choice $N_t=300$ is highlighted with a red cross, and the rest of parameters coincide with Fig.~\ref{fig:gamK}(b). See Appendix~\ref{ap:numerical} for details on the numerical calculation.}
 \label{fig:HHGpulse_tempDiv}
\end{figure}

\begin{figure}[tbp]
\centering
  \includegraphics[width=\linewidth]{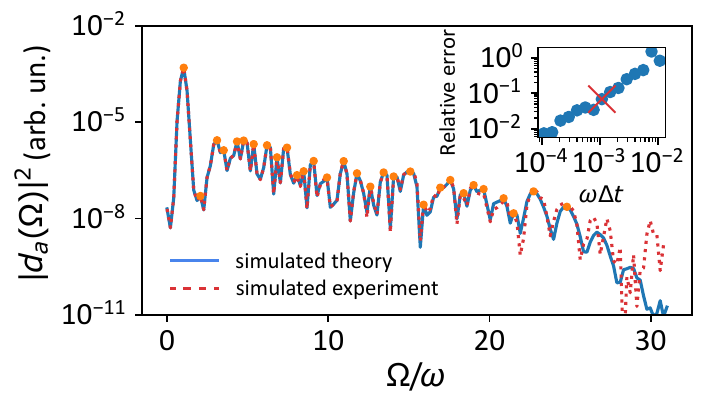}
      \caption{mission yield extracted from the discrete Fourier transform where the position of the temporal intervals randomly deviate from a uniform distribution with a standard deviation, $\omega \Delta t\sim 10^{-3}$ (red dotted line), as compared to the perfect spectrum (blue line). Orange dots indicate the positions of local maxima before the cutoff energy. In the inset, we show the relative error of these local maxima for increasing values of standard deviation, $\Delta t$, where the choice illustrated before is highlighted with a red cross. The rest of parameters coincide with Fig.~\ref{fig:gamK}(b). }
 \label{fig:HHGpulse_temp}
\end{figure}

\begin{figure}[tbp]
\centering
  \includegraphics[width=\linewidth]{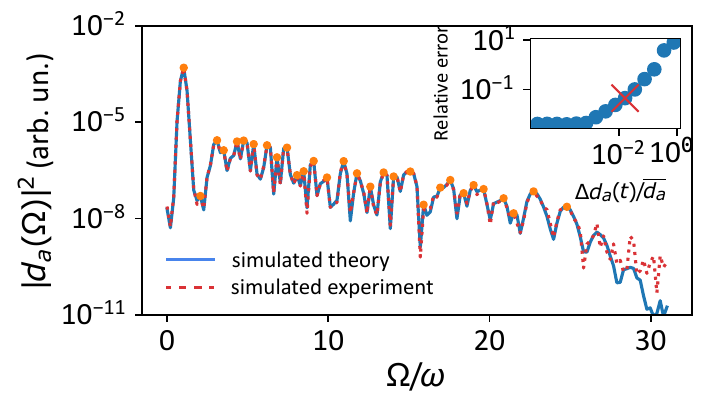}
      \caption{Emission yield extracted from the discrete Fourier transform of the temporal dipole acceleration, whose values are randomly distorted with standard deviation, $\Delta d_a(t)/\bar d_a=0.02$ (red dotted line), as compared to the perfect spectrum (blue line). Orange dots indicate the positions of local maxima before the cutoff energy. In the inset, we show the relative error of these local maxima for increasing values of standard deviation,  $\Delta d_a(t)$, where the choice illustrated before is highlighted with a red cross. The rest of parameters coincide with Fig.~\ref{fig:gamK}(b). }
 \label{fig:HHGpulse_Nmeas}
\end{figure}

\begin{figure}[tbp]
\centering
  \includegraphics[width=\linewidth]{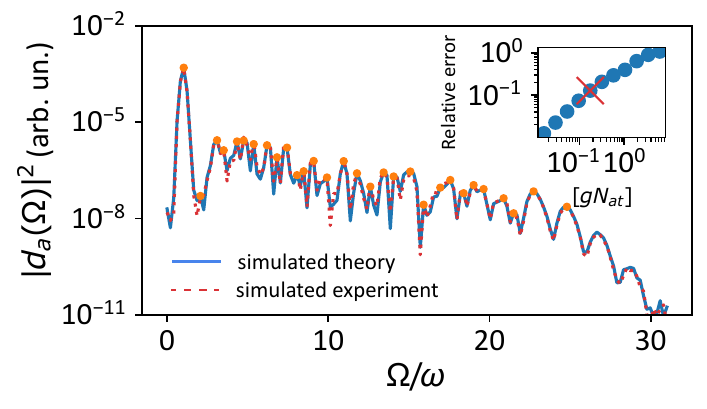}
      \caption{ Emission yield associated to an atomic cloud with interatomic interaction strength, $[gN\stx{at}]=0.2$ (red dotted line), as compared to the noninteracting case (blue line). Orange dots indicate the positions of local maxima before the cutoff energy. In the inset, we show the relative error of these local maxima for increasing values of interacting strength, where the choice illustrated before is highlighted with a red cross. The rest of parameters coincide with Fig.~\ref{fig:gamK}(b). }
 \label{fig:HHGpulse_int}
\end{figure}

The reconstruction of the dipole acceleration is also affected by experimental imperfections and limitations that need to be accounted for. Here, we numerically benchmark the main sources of error for the previous configuration $K_0=6.31$, $\gamma=0.81$ and $n_c=6$, where a larger conversion efficiency is expected [crossed marker in Fig.~\ref{fig:gamK}(a)].
To quantify the effect of these imperfections, we calculate the relative error in the determination of the local maxima of the spectra for frequencies below the cutoff frequency, which we highlight with orange dotted markers along Figs.~\ref{fig:HHGpulse_tempDiv}-\ref{fig:HHGpulse_int}.

In the experiment, the time-dependent dipole acceleration is measured at a finite set of times, $d_a(t_n)_{n=1\ldots N_t}$. The accuracy of the discrete Fourier transform
\begin{equation}
    d_a(\Omega)=\frac{1}{N_t}\sum_{n=1}^{N_t} d_a(t_n) e^{i\Omega t_n}\,,
\end{equation}
then depends on the number of time points used in the reconstruction, that we consider to be uniformly distributed along the duration of the pulse.
In the inset of Fig.~\ref{fig:HHGpulse_tempDiv}, we calculate the relative error in the local maxima of the emission yield for different values of temporal divisions. We observe that a moderate number of time intervals $N_t\sim 300$ provide an error of $\sim 1\%$ that is enough to resolve the cutoff energy (see red dashed line in Fig.~\ref{fig:HHGpulse_tempDiv}). 

Experimental errors in the time points used to measure, $\Delta t_n=|t\stx{meas}-t_n|$, would also manifest in the reconstructed yield. In the inset of Fig.~\ref{fig:HHGpulse_temp}, we calculate the relative error in the local maxima of the emission yield for different values of noise with standard deviation $\Delta t$ around the time intervals $t_n$. We observe that moderate values $\omega\cdot \Delta t\sim  10^{-3}$ provide an overall relative error $\sim 10\%$ in the local maxima that is enough to resolve the cutoff energy (see red dotted line in Fig.~\ref{fig:HHGpulse_temp}). Expressed in experimental units, this translates into a correct control of time in the order of $\sim 10\,\mu$s.
Overall, we see that the highest harmonics are the most sensitive ones to errors in the timing and number of time intervals, as they need to capture the rapidly oscillating response of the emission spectrum.

To quantify the time-dependent dipole acceleration from absorption measurements one also needs to characterize the needed accuracy on the determination of $d_a(t)$.
In the inset of Fig.~\ref{fig:HHGpulse_Nmeas}, we calculate the relative error in the local maxima of the emission yield for different values of relative error in the measurement of the time-dependent dipole acceleration, $\Delta d_a(t)/\overline{d_a}$, where $\overline{d_a}=\int_0^T |d_a(t)|/T$ is the temporal average throughout the pulse. We observe that $\Delta d_a(t)/\overline{d_a} \sim 1\%$ provides an accuracy in the emission yield that is enough to resolve the cutoff energy (see red dotted line in Fig.~\ref{fig:HHGpulse_Nmeas}). We also observe that the highest harmonics, which have the smallest intensity, are the most affected by a limited resolution. Following the TOFM approach to access $d_a(t)$, typical imaging lengths, $0.1-1.0$ mm, associated to expansion times $\tau\sim 10-100$ ms~\cite{bergschneiderSpinresolved2018}, can provide the needed accuracy of $1\%$ on the retrieved velocities for an atomic cloud that has an initial width of $1-10\,\mu$m before its ballistic expansion. Also, an error in the measured density of order $\Delta |\psi|^2\sim 10^{-4}$ can be tolerated to reach the desired relative error of $1\%$ in the extracted time-dependent dipole (see Appendix C). As the atomic cloud spreads over its different velocity components, this results in $N\stx{ev}\sim 10^6$ single-atom events to reach the needed accuracy on velocities and atomic densities during the TOFMs used to extract each instantaneous dipole acceleration.

Considering all these demands, one can estimate the overall experimental time needed to reconstruct the time-dependent dipole acceleration. For an atomic cloud with $N\stx{at}\sim 10^4$ atoms~\cite{senaratneQuantum2018} and an estimated running time of $\tau\sim 1$ s per experiment, collecting enough statistics translates into a reasonable running time $\tau \cdot N\stx{ev}\cdot N_t/N\stx{at} \sim 10$ h. The major demand for resources comes from the repetition rate of the experiment and the required number of events needed to resolve the small variations in the absorption images. First experimental realizations are then specially suitable for the simulation of the first harmonics in the plateau region, as they are associated with larger conversion efficiencies, reducing the number of independent measurements that are required and, with that, the overall experimental time.

As a final remark, we have seen that thousands of atoms in an atomic cloud are used to simulate the state of a single electron. Atoms do, however, experience interactions at distances characterized by the scattering length, $a_s$. In a mean-field approach, the effect of these interactions can be described by the Gross–Pitaevskii effective Hamiltonian that depends on the atomic density at each point of space~\cite{lewensteinUltracold2012}
\begin{equation}
\label{eq:hamiltonianGross}
    \hat H\stx{int}(\rr,t)=\hat H(\rr,t) + gN\stx{at} |\psi(\rr,t)|^2\,,
\end{equation}
where, $g=4\pi\hbar^2 a_s/m$. In the inset of Fig.~\ref{fig:HHGpulse_int}, we calculate the relative error in the retrieved spectrum for increasing values of interaction, observing an error below $10\,\%$ for $[gN\stx{at}]\sim 0.2$.
As compared to the previously discussed sources of error, we observe in Fig.~\ref{fig:HHGpulse_int} that the presence of interactions distort the entire spectrum, and not only the largest harmonics.
Expressed in experimental parameters, for a gas of $N\stx{at} \sim 10^4$ atoms of $^{84}$Sr, this value of interaction corresponds to $a_s= r_0 [gN\stx{at}]/(4\pi N\stx{at} [r_0])\sim 10^{-12}$ m, which is three orders of magnitude shorter than the scattering length of the experiment in Ref.~\cite{senaratneQuantum2018}. Using magnetic atoms such as Li, one could further rely on Feshbach resonances to engineer the required small values of scattering length~\cite{chin10a}.

\section{Outlook}
\label{sec:discussion}
In this work, we have shown that the HHG emission yield can be simulated in current analog experiments using atomic clouds. We have characterized the response of the simulator to elliptic potentials and short pulses, as well us the main sources of errors for experimental implementations. Based on our calculations, we see that simulating attosecond dynamics by means of HHG is indeed possible with current analog simulators. They also offer a richer tunability on both the simulated incoming fields and the effective nuclear potential, where
spatially modulated light can engineer potentials~\cite{Nogrette2014,bluvsteinQuantum2022} that mimic the long-range coulomb attraction that ionized electrons experience during their excursion time. 
The flexibility of this approach offers several opportunities for the simulation of utltrafast processes:
\begin{itemize}
\item As non-classical light sources become intense enough to drive the process of HHG~\cite{gorlachHighharmonic2023}, it reveals the limitations of semi-classical approximations~\cite{stammer2023limitations}, such as simple man's model that treats the driving field purely classically. This implies new challenges for semi-classical methods to accurately predict the HHG spectrum~\cite{gorlachHighharmonic2023} or the harmonic coherence properties~\cite{stammer2023role} for such cases. Interestingly, even if photons do not have a counterpart in this simulator, the response of the system to non-classical light (e.g. squeezed pulses) can be approximated by an effective photon-statistics force~\cite{eventzurPhotonstatistics2023} that can potentially be included in the simulator thanks to the high tunability of the effective incoming field. 
\item Further extensions of this platform would focus on studying multielectronic processes. For example, using shaken potentials in the ATI regime and the refined control provided by optical tweezers, each electron can be mapped to an individual fermionic atom in the trap. Extended atom-atom repulsive interactions provided by e.g., magnetic dipoles~\cite{baierExtended2016}, Rydberg dressing~\cite{guardado-sanchezQuench2021,sibalicRydberg2018} or mediating particles~\cite{arguello-luengoEngineering2021a, Chang2018}, allow one to directly tailor Coulomb corrections for a tunable degree of electronic repulsion. Going beyond the single-active electron approximation for HHG, this offers the opportunity to simulate processes such as NSDI, where the electron-electron correlation plays a pivotal role and needs to be included in the theoretical models to satisfactorily describe the experimental observations~\cite{weber_correlated_2000,moshammer_momentum_2000,weber_recoilion_2000,rudenko_correlated_2007,staudte_binary_2007}. 
\item In this direction, the creation of tweezer arrays of controllable geometries can also assist with the understanding of rescattering processes that take place in the crystalline structure of materials~\cite{osika_wannier-bloch_2017,parks_wannier_2020}. In this context, conventional theoretical analyses of HHG processes in solid-state media result in quasicontinuum harmonic spectra, lacking the clear presence of harmonic peaks that would be expected under ideal conditions. To make these harmonics distinctly visible, exceedingly small dephasing times (typically on the order of $\sim\!1$ fs) are numerically employed~\cite{vampa_theoretical_2014}, which may conflict with experimental observations~\cite{becker_femtosecond_1988,portella_kspace_1992} where the observed dephasing times are in the range of 15-50 fs. While recent theoretical studies have suggested that achieving numerical convergence is crucial to obtain well-resolved harmonic spectra without requiring extremely low dephasing times~\cite{kolesik_numerical_2023}, others have tried to explain this phenomenon by considering the effect of propagation~\cite{kilen_propagation_2020}, or destructive interference between spatially long trajectories caused by the spatial distribution of laser fields ~\cite{brownRealSpace2022}. In this regard, using a cleaner environment compared to real materials can offer a higher temporal and spatial resolution that can help discern the relevance of these interference effects. Furthermore, it can also be utilized to study the connection between the delocalized recombination of electrons~\cite{osika_wannier-bloch_2017,parks_wannier_2020,yue_imperfect_2020} and the reduced dependence of the harmonic yield with the driving field's ellipticity~\cite{ghimire_observation_2011}. 
\end{itemize}
Overall, the tunability and accessibility of these experiments can be of significant value as a complementary tool to benchmark recently accessed experimental regimes, and stimulate the development of novel numerical techniques and theoretical models, helping to reach a more profound understanding of the electronic response of atoms and materials to ultrafast and intense laser fields. 

\section*{Acknowledgements}
We acknowledge Anna Dardia, Peter Dotti, Toshihiko Shimasaki, and Yifei Bai for helpful discussions
on the experimental implementation. The ICFO group
acknowledges support from: the European Research Council (ERC) Advanced Grant (AdG) “NOvel Quantum
simulators—connectIng Areas” (NOQIA) (Grant agreement No. 833801); Plan National STAMEENA PID2022-
139099NB-I00 project funded by MCIN/AEI/10.13039/
501100011033 and by the “European Union NextGenerationEU/PRTR” (PRTR-C17.I1), FPI); the Ministerio de
Ciencia y Innovation Agencia Estatal de Investigaciones
(PGC2018-097027-B-I00/10.13039/501100011033, CEX
2019-000910-S/10.13039/501100011033, Plan National
FIDEUA PID2019-106901GB-I00, FPI, QUANTERA
MAQS PCI2019-111828-2, QUANTERA DYNAMITE
PCI2022-132919 (QuantERA II Programme co-funded
by European Union’s Horizon 2020 program under
Grant Agreement No 101017733), Ministry of Economic Affairs and Digital Transformation of the Spanish Government through the QUANTUM ENIA project
call – Quantum Spain project, and by the European Union through the Recovery, Transformation, and
Resilience Plan – NextGenerationEU within the framework of the Digital Spain 2026 Agenda; Proyectos de
I+D+I “Retos Colaboraci” QUSPIN RTC2019-007196-
7); MICIIN, with funding from the European Union
(EU) NextGenerationEU (Grant No. PRTR-C17.I1) and
by the Generalitat de Catalunya; the Fundació Cellex;
the Fundació Mir-Puig; the Generalitat de Catalunya
(European Social Fund FEDER and CERCA programs, AGAUR Grant No. 2021 SGR 01452, QuantumCAT
U16-011424, cofunded by the European Regional Development Fund (ERDF) Operational Program of Catalonia 2014–2020); the Barcelona Supercomputing Center MareNostrum (Grant No. FI-2023-1-0013); the EU
(PASQuanS2.1, 101113690); the EU Horizon 2020 FETOpen “Optical Topologic Logic” (OPTOlogic) program
(Grant No. 899794); and the EU Horizon Europe Program
(Grant Agreement No. 101080086—NeQST), National
Science Centre, and ICFO Internal “QuantumGaudi”
project. European Union’s Horizon 2020 program under
the Marie Sklodowska-Curie grant agreement No 847648;
“La Caixa” Junior Leaders fellowships, La Caixa” Foundation (ID 100010434): CF/BQ/PR23/11980043. J.R.-D.
acknowledges funding from the Secretaria d’Universitats
i Recerca del Departament d’Empresa i Coneixement
de la Generalitat de Catalunya, the European Social
Fund (L’FSE inverteix en el teu futur)–FEDER, and
the ERC AdG Certification of quantum technologies
(Grant agreement No. 834266). P.S. acknowledges funding from the EU Horizon 2020 research and innovation
program, under the Marie Skłodowska-Curie Grant Agreement No. 847517. A.S.M. acknowledges funding support
from the EU Horizon 2020 research and innovation program under the Marie Skłodowska-Curie Grant Agreement SSFI No. 887153. D.M.W. acknowledges support
from the Air Force Office of Scientific Research (Grant
No. FA9550-20-1-0240) and the University of California Santa Barbara NSF Quantum Foundry funded via the
“Enabling Quantum Leap: Convergent Accelerated Discovery Foundries for Quantum Materials Science, Engineering and Information” (Q-AMASE-i) program under
Grant No. DMR1906325. M.F.C. acknowledges financial
support from the Guangdong Province Science and Technology Major Project (“Future Functional Materials under
Extreme Conditions”—Grant No. 2021B0301030005) and
the Guangdong Natural Science Foundation (General Program Project No. 2023A1515010871). Views and opinions
expressed are, however, those of the author(s) only and do
not necessarily reflect those of the European Union, European Commission, European Climate, Infrastructure and
Environment Executive Agency (CINEA), or any other
granting authority. Neither the European Union nor any
granting authority can be held responsible for them.
%

\clearpage
\appendix
\renewcommand\thefigure{C\arabic{figure}}  
\renewcommand\appendixname{}
\setcounter{figure}{0} 
\setcounter{page}{1} 
\newpage
\section*{APPENDIX}

\section{The Kramers-Henneberger correspondence}
\label{ap:Kramers-Henneberger correspondance}
One approach to simulate an oscillatory force on the simulating atom consists of shaking the trap potential. For simplicity, let us start considering a one-dimensional system trapped in the effective potential $V_n(x)$. When the potential is shaken with amplitude $L$ and frequency $\omega$, the effective atomic Hamiltonian is
\begin{equation}
\label{eq:H1}
   \hat H\stx{lab}(x,t)=\frac{p^2}{2m}+V_n[x+L\cos(\omega t)] \,,
\end{equation}
with momentum $p$. To see the Kramers-Henneberger (KH) correspondence to an effective dipole force, one can follow the derivation in Ref.~\cite{dreseUltracold1997}. After the unitary displacement, $U_1=\exp\co{-iL\cos(\omega t)p/\hbar+imL^2\omega\sin(2\omega t)/(8\hbar)}$, the rotated Hamiltonian, $H'=U_1H\stx{lab}U_1^\dagger+i\hbar (\partial_tU_1)U_1^\dagger$ writes as
\begin{equation}
  \hat  H'(x,t)=\frac{\co{p-mL\sin(\omega t)}^2}{2m}+V_n(x)-\frac{mL^2\omega^2}{4}.
\end{equation}

Performing now a momentum displacement $U_2=\exp\co{-imL\omega\sin(\omega t)x/\hbar}$ one finds the KH frame
\begin{equation}
\label{eq:H2}
 \hat  H\stx{KH}(x,t)=\frac{p^2}{2m}+V_n(x)+mL\omega^2\cos(\omega t)x-\frac{mL^2\omega^2}{4},
\end{equation}
which, up to the energy shift, $mL^2\omega^2/4$, corresponds to the static Hamiltonian with an additional effective oscillating force of strength, $F_0=mL\omega^2$. In this rotated frame, the initial state $\ket{\psi_{0,\text{KH}}} = U_2 U_1 \ket{\psi_{0}}$, where $\ket{\psi_{0}}$ solves \eqref{eq:H1}, should correspond to the bound-state of the trap, $\ket{0}$, which is also the state that can be easily prepared in the lab-frame, $\ket{\psi_0}=\ket{0}$. The validity of this preparation is then dictated by
\begin{equation}
\label{eq:mismatch}
    \alpha|_{t=0}\equiv|\braket{\psi_{0,\text{KH}}}{ \psi_0}|=|\bkev{0}{U_2^\dagger U_1^\dagger }{0}|\sim e^{-1/\gamma^2}\,,
\end{equation}
which makes the shaking approach specially suitable for the multiphoton ionization regime, $\gamma\gg 1$, where $\alpha \sim 1$, but challenging in the HHG regime $\gamma\lesssim 1$. 

Focusing on pulses with a finite number of cycles, the field is gradually introduced by the carrier envelope, which prevents the direct mismatch derived in Eq.~\eqref{eq:mismatch}. Still, one should observe that the atomic cloud cannot follow the shaken trap if the applied gradient of the potential, $\partial_x V_n(x)$, is weaker than the simulated inertial force of interest, $F_0$. For the Gaussian trap in Eq.~\eqref{eq:nuclearPot}, this semiclassical condition writes as $V_0 \gg F_0r_0$, which challenges the experimental realization of the region $I_p\sim F_0r_0$ introduced in Eq.~\eqref{eq:tunnelCrit}, where the largest conversion efficiency of HHG is expected.
This motivates the use of real oscillatory potentials to simulate this scenario, as we introduce in the main text.

\section{Connection to atomic units}
\label{ap:units}
Relevant chemical values are often expressed in atomic units (that we denote as $[[\cdot]]$): $[[\hbar]]= [[m_e]]= [[k_e]]= [[e]]\equiv 1$. Conveniently, the Bohr radius ($a_B$) and Hartree energy, associated with the width and energy of the ground state of atomic Hydrogen, write as
$$
\co{[a_B]}=\co{\co{ \frac{\hbar^2}{e^2m_ek_e}}}\equiv 1 \,; \quad \co{[\text{Hartree}]}= \co{\co{ \frac{e^4k_e^2m_e}{\hbar^2}}}\equiv 1\,.
$$

The Hamiltonian of our simulator is of a different form, as the nuclear potential $e^2k_e/r$ is replaced by a Gaussian profile, $V_0\exp\co{-x^2/(2 r_0)^2}\approx V_0\co{1-x^2/(2r_0^2)}$.
The natural units of the system now write as 
$$[\hbar]= [m]=[e]=[V_0 r_0]\equiv 1\,.$$
Thus, the characteristic energy ($E_0$) and length scales ($a_0$) of the system are
$$
[E_0]=\co{m\pa{\frac{V_0 r_0}{\hbar}}^2}\equiv 1\,,
$$
$$
[a_0]=\co{\frac{\hbar^2}{mV_0 r_0}}\equiv 1\,.
$$

In these units, the attoscience Hamiltonian~\eqref{eq:hamiltonian} writes to the lowest order (up to a constant energy shift)
\begin{equation*}
    \hat H \approx -\frac{\partial^2_{[x]}}{2} + \frac{[x]^2}{2[r_0]^3} + [F_0]\sin([\omega]t)\,[x]\,,
\end{equation*}
which also matches the lowest order expansion of the soft-core potential $V\stx{sc}(x,r_0)$ for $[V_0r_0]=1$. Following this approach, the parameters of the simulator expressed in these units coincide with the same dimensionless value as the magnitudes of the real-life experiment expressed in atomic units. The connection with the other parameters of the model is finally derived as:
\begin{align*}
    [\omega]&= \omega \cdot \hbar/E_0\,,\\
    [F_0]&=F_0\cdot a_0/E_0\,,\\
    [I_p]&=I_p/E_0 \,, \\
    [c]&=c\cdot\hbar/(a_0E_0)\,.
\end{align*}

\section{Numerical methods}
\label{ap:numerical}
In the simulation of Fig.~\ref{fig:HHGpulse_field} we consider the atomic wavefunction extended over a length $x\stx{max}=180\,r_0$, that is divided into 3000 discrete points. To prevent reflections at the edges, we apply a smooth attenuation mask of the form $\cos^{1/8}$ along the last $3r_0$ length of the array~\cite{krauseCalculation1992}. 

To solve the time-dependent Schr\"odinger equation, we calculate the evolution of an initial ground-state of the trapping potential, $\ket{\psi(0)}=\ket{0}$, under the time-dependent Hamiltonian~\eqref{eq:H2}.
This evolution is Trotterized as~\cite{feitSolution1982}
\begin{equation}
\begin{split}
        \ket{\psi(t)}&=e^{-\frac{i}{\hbar}\int_0^t H(s)ds}|0\rangle \\
&= \prod_{n=0}^{\fl{t/\tau_0}} \pa{e^{-\frac{i \tau_0}{\hbar} H_x(n \tau)} e^{-\frac{i  \tau_0}{\hbar}H_p}} |0\rangle +\mathcal{O}(\tau^2) \,,
\end{split}
\end{equation}
where the operators $H_x(t)=V(x)+F_0\cos(\omega t)x$ and $H_p=p^2/(2m)$ are diagonal in real and momentum space respectively. This simplifies their exponentiation by appropriately Fourier transforming the state from the real to the momentum space, where either of the trotterized evolutions is diagonal. Along the Figures of this work, we considered $\tau_0\omega=0.01$. 

\begin{figure}[tbp]
\centering  \includegraphics[width=\linewidth]{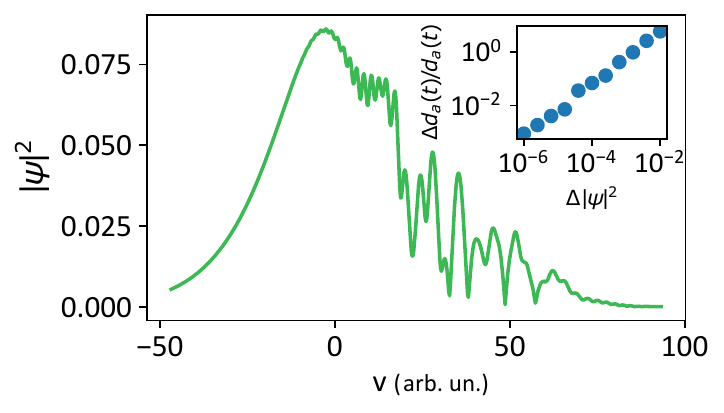}
      \caption{ Velocity components of the atomic cloud, $|\psi(v)|^2$ after half of an incoming 6-cycle pulse is applied for the same parameters as in Fig.~\ref{fig:gamK}(b). The inset shows the relative error on the reconstructed dipole acceleration, $d_a(t=T/2)$ as one introduces Gaussian noise of standard deviation $\Delta |\psi|^2$ on the velocity components of the main Figure. The markers indicate the average over 10 random realizations.}
 \label{fig:psiV}
\end{figure}

In Fig.~\ref{fig:psiV}, we show the velocity components of the atomic cloud (corresponding to the squared wavefunction in momentum space) at a given instant of the driving. This curve can be measured in the experiment through TOFM, where the momentum is mapped to different positions of the atomic cloud after a ballistic expansion of the gas. In the inset, we show that the relative error of the extracted dipole acceleration scales polynomially with the noise on the atomic density detection, providing an error smaller than $1\%$ for $\Delta |\psi|^2\sim 10^{-4}$. This indicates a desired accuracy for experimental fluorescence measurements.

\end{document}